\documentstyle[11pt,epic,eepic]{article}

\title{\LARGE\bf
Retarded Electromagnetic Interaction and Origin of Non-linear 
Phenomena in Ferroelectrics and Ferromagnetics as well as Optics}
\author{Mei  Xiaochun\\
Department of Physics, Fuzhou University, Fuzhou, 350025, China\\
E-mail: fzbgk@pub3.fz.fj.cn}
\date{February 1, 2003}

\begin{document}
\maketitle

\begin{abstract}
The non-linear relations between polarization strength and electric field strength for ferroelectrics, as well as magnetization strength and
magnetic field strength for ferromagnetics, can be achieved by introducing the retarded electromagnetic interactions. The electric and 
magnetic hysteretic loops in medium frequency alternating electromagnetic fields can be described well. The basic phenomenological formula
in current non-linear optics can also be obtained in the high frequency fields. The results show that all of these non-linear phenomena
come from the retarded electromagnetic interaction.
\end{abstract}
\par

\par
According to classical electromagnetic theory, the Maxwell equations in medium are
\begin{equation}
\nabla\times\vec{E}=-{{\partial\vec{B}}\over{\partial{t}}}~~~~~~~~~~~~~~~~~~\nabla\times\vec{H}=\vec{J}_f+{{\partial\vec{D}}\over{\partial{t}}}
\end{equation}
\begin{equation}
\nabla\cdot\vec{D}=\rho_f~~~~~~~~~~~~~~~~~~~~~~~~~\nabla\cdot\vec{B}=0
\end{equation}
The experiments show that for general non-ferroelectrics and non-ferromagnetics in weak electromagnetic fields, the relations between electric polarization strength $\vec{P}$ and $\vec{E}$, as well as magnetization strength $\vec{M}$ and $\vec{H}$ are linear, connected by the so-called constructive equations
\begin{equation}
\vec{D}=\varepsilon_0\varepsilon\vec{E}=\varepsilon_0\vec{E}+\vec{P}~~~~~~~~~~~~~\vec{B}=\mu_0\mu\vec{H}=\mu_0(\vec{H}+\vec{M})
\end{equation}
\begin{equation}
\vec{P}=\varepsilon_0\chi_e\vec{E}~~~~~~~~~~~~~~~~~\vec{M}=\mu_0\chi_m\vec{H}
\end{equation}
Here $\chi_e$ is polarizability and $\chi_m$ is magnetic susceptibility individually with $\varepsilon=1+\chi_e$, $\mu=1+\chi_m$. For non-isotropy electric and magnetic mediums, the relations are still linear with simple forms
\begin{equation}
\vec{P}=\varepsilon_0\vec{\vec{\chi}}_e:\vec{E}~~~~~~~~~~~~~\vec{M}=\mu_0\vec{\vec{\chi}}_m:\vec{H}
\end{equation}
Here $\vec{\vec{\chi}}_e$ is polarizability tensor and $\vec{\vec{\chi}}_m$ is magnetic susceptibility tensor.
\par
However, the experiments also show that for ferroelectrics and ferromagnetics, as well as in the strong fields, the relations between $\vec{P}$ and $\vec{E}$, as well as $\vec{M}$ and $\vec{H}$ are non-linear with great complexity. Their relations can¡¯t be deduced from theory at present. We now get them only by phenomenological hypothesis. In the polarization and magnetization processes of ferroelectrics and ferromagnetics, there exist hysteretic phenomena. In the alternating electromagnetic fields of low frequency (for example 0.1Hz), the magnetic hysteretic loops and magnetization curve are shown in Figure 1. For ferromagnetics, $M$ is much big than $H$, so the shape of $M\sim{H}$ curve is similar to the shape of $B\sim{H}$ curve. The magnetic hysteretic loops are near rectangle. In the figure, the maximum values $M_0$ and $H_0$ are at the same point. By fixing the frequency of magnetic field and changing $H_0$, we can get a series of loops. Connecting the vertex of each loop, we get a curve, i.e., magnetization curve.
\par
When the frequency becomes high, the hysteretic loops become smooth ellipses. When the frequency becomes higher, the ellipses become near circular. Figures 2 and 3 are the magnetic hysteretic loops of nickel-zinc ferrite in the magnetic fields with frequencies 1500Hz and 500Hz individually. 
\par
When the frequencies are high enough, the maximum values $B_0$ and $H_0$ are not at the same point. So we can¡¯t get the magnetization curves by the same method shown in Figure 1. By considering continuity of magnetization processes, in medium frequency fields, we can estimate the basic sharps of magnetization curves below when the frequency and the maximum value of magnetic field are fixed. Because magnetization curves must reach the maximum value of magnetic field but can¡¯t pass over the value, the magnetization curves can take three forms shown in Figure 3, described in broken lines a, b and c. The curves a and b are unclosed at first several cycles. It needs some time for them to become stationary loops. Curve c can smoothly enter stationary loop at its first cycle before it reaches the maximum point of magnetic field. In fact, the situation is the same for magnetization curve in low frequency fields. The Loops shown in Figure 1 are only stationary ones. At the beginning of magnetization processes, they should also be unclosed. The forms of hysteretic loops and polarization curves ferroelectrics are the same as ferromagnetics.

\begin{figure}[h]
\begin{center}
\begin{picture}(200,240)(-100,-120)
\put(-100,-120){\tiny \grid(200,240)(50,20)[-100,-120]}
\dottedline{1.4}(-100,-100)(-88.2,-98)(-83.8,-96)(-76.5,-94)(-69.1,-92)(-58.8,-84.6)(-52.9,-78.1)(-44.1,-64.5)(-39.7,-55)
(-34.6,-42.1)(-27.9,-27.2)(-19.1,-12.4)(-10.3,-6)(-5.2,-3)(0,0)(5.2,2.4)(11,4.7)(13.2,5.9)(19.9,11.8)(23.5,17.8)(27.9,26)(33.8,40.8)
(41.9,58.6)(51.5,76.9)(62.5,88.2)(75,95)(86.8,99.4)(97,101.8)
\drawline(-100,-100)(-86,-98.8)(-52.9,-91.7)(-39.7,-88.8)(-25,-84)(-6.6,-76.3)(1.5,-72.2)(19.1,-59.2)(25.7,-53.9)(32.4,-43.8)(42.7,-21.9)
(46.3,-2.98)(47.1,20.7)(48.5,40.8)(53.7,63.3)(65.4,82.3)(72.8,88)(82.4,96.5)(97.1,101.8)(80.2,99.4)(54.4,94.1)(47.8,91.7)
(31.6,86.4)(21.3,83.4)(13.2,80.5)(4.4,76.3)(-2.9,73.4)(-16.9,62.1)(-22.8,54.4)(-29.4,43.8)(-34.6,31.4)(-39.7,-6.5)(-41.2,-20.7)(-45.6,-43.8)
(-51.5,-62.1)(-58.8,-75.7)(-69.9,-85.8)(-78.7,-91.7)(-89.7,-96)(-100,-100)

\drawline(-39.7,-55)(-30.9,-53.9)(-19.1,-50.3)(-10.3,-46.2)(0.74,-41)(9.6,-32)(22.8,-18.3)(33.8,2.3)(38.2,28.4)(39,45.)(40.4,55)
(27.2,53.3)(19.1,50.9)(10.3,46.2)(-2.9,39.1)(-14,27.2)(-23.5,13)(-32.4,-8.9)(-36.8,-35.5)(-39,-46.2)(-39.7,-55)
\drawline(-17.7,-11.2)(-10.3,-12.5)(-3.7,-11.5)(5.2,-7.5)(12.5,-1)(16.2,8.9)(6.6,7.5)(0.7,6.5)(-5.2,4)(-14.7,-3)(-17.7,-11.2)
\end{picture}
\end{center}
\caption[]{}
\end{figure}

\begin{figure}[h]
\begin{center}
\begin{picture}(200,240)(-100,-120)
\put(-100,-120){\tiny \grid(200,240)(50,20)[-100,-120]}
\drawline(-54,-116.2)(-53.5,-116.2)(-50,-120.1)(-40,-121.5)(-36.2,-120.1)(-29.23,-117.7)(-20.8,-113.9)(-10,-106.2)(-1.5,-95.4)(3,-83.1)
(14,-53.9)(30,1.5)(48.9,73.1)(57.7,110.8)(58.5,118.5)(56.2,123.1)(50.8,127.7)(43.9,126.9)(33.9,122.2)(25.4,115.4)(15.4,102.3)(4,81.5)
(1.5,73.9)(-12.3,33.9)(-22.3,0.8)(-41.5,-66.2)(-48.5,-91.5)
(-52.5,-99.2)(-53.5,-111.5)(-54,-116.2)
\end{picture}
\end{center}
\caption[]{}
\end{figure}

\begin{figure}[h]
\begin{center}
\begin{picture}(200,240)(-100,-120)
\put(-100,-120){\tiny \grid(200,240)(50,20)[-100,-120]}
\drawline(0.9,-76.4)(10.91,-66.4)(26.4,-48.2)(38.2,-34.6)(51.8,-15.5)(60,-1.8)(66.4,12.7)(71.8,24.6)(78.2,43.6)(81.8,64.6)(80,88.2)(76.4,95.5)
(70.9,100.9)(61.8,105.5)(49.1,105.5)(38.2,101.8)(26.4,97.3)(5.5,81.8)(-13.6,64.6)(-32.7,42.7)(-51.8,14.6)(-68.2,-16.4)(-76.4,-48.2)
(-79.1,-71.8)(-73.6,-90)(-69,-94.6)(-62.8,-98.2)(-54.6,-100)(-41.8,-99.1)(-33.6,-95.5)(-25.5,-92.7)(0.9,-76.4)
\dottedline{1.4}(0,1.8)(10.9,7.3)(29.1,18.2)(44.6,30)(54.6,39.1)(67.3,52.7)(77.3,71.8)(84.6,97.3)(80.9,110)(74.6,118.2)(70,120.9)(57.3,125.5)
(49.9,126.4)
\dottedline{1.4}(0,1.82)(20,2.73)(41.8,4.6)(59.1,9.1)(75.5,21)(82,35.5)(81.8,44)(79.1,49.1)(72.7,53.6)(70.9,54.6)(57.3,60)(40,64.5)
\dottedline{1.4}(0,1.82)(19.1,5.5)(36.7,12.7)(54.6,24.6)(70,42.7)(79.1,66.4)(81.8,81.8)
\end{picture}
\end{center}
\caption[]{}
\end{figure}

\begin{figure}[h]
\begin{center}
\begin{picture}(200,200)(-100,-100)
\put(-100,-100){\tiny \grid(200,200)(50,50)[-100,-100]}
\drawline(-100,-100)(-98,-100)(-94,-100)(-87,-99)(-77,-98)(-64,-94)(-50,-88)(-34,-78)(-17,-65)(0,-50)
(17,-31)(34,-10)(50,12)(64,35)(77,57)(87,75)(94,88)(98,97)(100,100)
\drawline(-100,-100)(-98,-97)(-94,-88)(-87,-75)(-77,-57)(-64,-35)(-50,-12)(-34,10)(-17,31)(0,50)(17,65)(34,78)(50,88)(64,94)(77,98)(87,99)(94,100)(98,100)(100,100)
\dottedline{1.4}(0,0)(17,10)(34,23)(50,38)(64,53)(77,68)(87,81)(94,91)(98,97)(100,100)
\end{picture}
\end{center}
\caption[]{}
\end{figure}
\clearpage

\par
It is still an unsolved problem how to deduce the function forms of hysteretic loops and polarization curves in theory now. What we have at present are the experimental formulas for ferromagnetics provided by L.Rayleigh in 1887 $^{(1)}$
\begin{equation}
B=\mu_0(\mu¡¯{H}+bH^2)~~~~~~~~~~~B=\mu_0(\mu¡¯+bH_0)H\pm\mu_0{b\over{2}}(H^2_0-H^2)
\end{equation}
The formulas are only suitable for weak fields. In the formula, $\mu¡¯$ is initial magnetic susceptibilities, $b$ is a constant. The positive sign is for the up-line of formula of magnetic hysteretic loop and the negative sign for down-line. Suppose alternating magnetic field is $H=H_0\sin\omega{t}$, let $\alpha=bH_0/\mu¡¯$, Eq.(6) can be written as 
\begin{equation}
B=\mu_0\mu¡¯{H}_0(\sin\omega{t}+\alpha{\sin}^2\omega{t})~~~~~~~~B=\mu_0\mu¡¯{H}_0[(1+\alpha)\sin\omega{t}\pm{{\alpha}\over{2}}{\cos}^2\omega{t}]
\end{equation}
Take $\alpha=1$, let $H_0$ as the unit of abscissa, $B_0/(2\mu_0\mu¡¯)$ as the unit of ordinate, we can get the magnetic hysteretic loop and magnetization curve shown in Fig.4. It can be seen that the loop is not an approximate rectangle, so it dose not consistent with Figure 1 in low frequency condition. It is not smooth and continuous at the maximum points, so it dose not consistent with Figure 2 and 3 in medium frequency fields. Meanwhile,. $B_0$ and $H_0$ are at the same point in figure 3, but the real situation is that they should not for the existence of phase difference between them.
\par
It is proved below that after retarded electromagnetic interaction is introduced, the non-linear relations between polarization strength and electric field strength, as well as magnetization strength and magnetic field strength can be obtained. Based on soft modal theory, the electric and magnetic hysteretic loops, as well as polarization and magnetization curves can be described well in the alternating electromagnetic fields of low and medium frequencies. The phenomenological formula used in non-linear optics can also be obtained by the same method in high frequency fields, meaning that the retarded electromagnetic interaction is the origin of all these non-linear phenomena.
\par
The polarization of ferroelectrics is discussed at first. According to the current theory, there exist small electrical domain structures in ferroelectrics under critical temperature. When the electrical polar moments in domains arrange in parallel, energy is lowest so that the system is most stable. Owing to the random distribution of polarization directions of electrical domains, when there exist no external fields, whole ferroelectrics are electrically neutral. When external electrical field is added, the polarization directions of domains are turned to the direction of electrical field, so that ferroelectrics show strong polarization. In alternating electrical field $E=E_0\sin\omega{t}$, the motion equation of an electrical polar moment is
\begin{equation}
{{d^2{r}}\over{dt^2}}+2\beta{{dr}\over{dt}}+\omega^2_0{r}={{qE_0}\over{m}}\sin\omega{t}
\end{equation}
The second item on the left side of equation represents the damping forces caused by collision and radiation and so on with $\beta>0$. The third item represents restoring force and, $\omega_0$ is the nature frequency of dipole and $q$ is dipole¡¯s charge. The solution of Eq.(8) is
\begin{equation}
r=A\sin(\omega{t}-\theta_0)+Ce^{-\beta{t}}\sin(\sqrt{\omega^2_0-\beta^2{t}}-\vartheta)
\end{equation}
\begin{equation}
A={{qE_0}\over{m\sqrt{(\omega^2_0-\omega^2)^2+(2\beta\omega)^2}}}~~~~~~~~~~\theta_0=tg^{-1}({{2\beta\omega}\over{\omega^2_0-\omega^2}})
\end{equation}
On the other hand, according to the soft modal theory of ferroelectrics $^{(2)}$, under critical temperature, the vibrations of lattices are frozen for the Coulombian forces and restoring forces of irons are just offset each other, when there exist no external electric fields. In this case, the dipoles would not vibrate with $\omega_0=0$.
\par
So let $\omega_0=0$ in Eq.(8), the solution of the i-th dipole¡¯s motion equation becomes
\begin{equation}
r_i=A_i\sin(\omega{t}-\theta_{i0})+B_i-{{C_i}\over{2\mu_i}}e^{-2\beta_i{t}}~~~~~~~~~~~~~~A_i={{q_i{E}_0}\over{m_i\omega^2\sqrt{1+(2\beta_i/\omega^2)^2}}}
\end{equation}
Here $B_i$ and $C_i$ are constants depending on initial condition. According to the theory of soft modal, when $t=0$ we have $r_i=r_{i0}$ and $v_i=v_{i0}=0$. For simplification, let $B_i=0$, by relation, $tg\theta_{i0}=-2\beta_i/\omega$, we get
\begin{equation}
r_{i0}=-A_i\sin\theta_{i0}~~~~~~~~~A_i={{q_i{E}_0\cos\theta_{i0}}\over{m_i\omega^2}}~~~~~~~C_i=-A_i\omega\cos\theta_{i0}
\end{equation}
So when $t\leq{0}$ the dipole has a stationery electrical polar moment $P_{i0}=q_i{r}_{i0}$. If $r_{i0}$ is positive, $\theta_{i0}$ is in the third or fourth quadrant. From $tg\theta_{i0}=-2\beta_i/\omega$, $\theta_{i0}$ is in the second or the fourth quadrant. So $\theta_{i0}$ should be in the fourth quadrant. Similarly, if $r_{i0}$ is negative, $\theta_{i0}$ should be in the second quadrant. From Eq.(11) and (12) we get
\begin{equation}
r_i=A_i\sin(\omega{t}-\theta_{i0})-{{A_i{\cos}^2\theta_{i0}}\over{\sin\theta_{i0}}}e^{-2\beta_i{t}}~~~~~~~~~~v_i=A_i\omega\cos(\omega{t}-\theta_{i0})-A_i\omega\cos\theta_{i0}{e}^{-2\beta_i{t}}
\end{equation}
\par
The retarded electromagnetic interaction is introduced below. Let $t¡¯$, $\vec{r}¡¯_i$, $\vec{v}¡¯_i$, $\vec{a}¡¯_i$ and $\dot{\vec{a}}¡¯_i$ represent retarded time, coordinate, velocity, acceleration and acceleration of acceleration, $t$, $\vec{r}_i$, $\vec{v}_i$, $\vec{a}_i$ and $\dot{\vec{a}}_i$ represent non-retarded time, coordinate, velocity, acceleration and acceleration of acceleration. A particle with charge $q_i$, velocity $\vec{v}¡¯_i$ and acceleration $\vec{a}¡¯_i$ at space point $\vec{r}¡¯_i(t¡¯)$ and time $t¡¯$ would cause retarded potentials at point $\vec{r}_i(t)$ and time $t$ as follows
\begin{equation}
\varphi¡¯_{ij}={{q_i}\over{4\pi\varepsilon_0(1-{{\vec{v}¡¯_i\cdot\vec{n}_{ij}}\over{c}})r¡¯_{ij}}}~~~~~~~~~~~\vec{A}¡¯_{ij}={{q_i\vec{v}¡¯_i}\over{4\pi\varepsilon_0{c}^2(1-{{\vec{v}¡¯_i\cdot\vec{n}_{ij}}\over{c}})r¡¯_{ij}}}
\end{equation}
In the formula $\vec{r}¡¯_{ij}(t,t¡¯)=\vec{r}_i(t)-\vec{r}¡¯_i(t¡¯)$, $t¡¯=t-r¡¯_{ij}(t¡¯)/c$, $r¡¯_{ij}=\mid\vec{r}¡¯_{ij}\mid$, $\vec{n}_{ij}=\vec{r}¡¯_{ij}/{r}¡¯_{ij}$. When $v¡¯_i<<c$, the retarded distance $r¡¯_{ij}(t¡¯)$ at retarded time $t¡¯$ can be replaced approximately by non-retarded distance $r_{ij}(t)$, i.e., we can let $t¡¯=t-r¡¯_{ij}(t¡¯)/c\rightarrow{t}-r_{ij}(t)/c$ and get
\begin{equation}
r¡¯_{ij}(t,t¡¯)=\mid\vec{r}_j(t)-\vec{r}¡¯_i(t¡¯)\mid=\mid\vec{r}_j(t)-\vec{r}¡¯_i[t-r¡¯_{ij}(t¡¯)/c]\mid\rightarrow\mid\vec{r}_j(t)-\vec{r}¡¯_i[t-r_{ij}(t)/c]\mid=r¡¯_{ij}(t)
\end{equation}
It is noted that $r¡¯_{ij}(t)\neq{r}_{ij}(t)$, for $r¡¯_{ij}(t)$ is the approximate retarded distance, but $r_{ij}(t)$ is not the retarded distance. In this case, we can develop retarded quantities into series in light of small quantity $r_{ij}/c$. By relation $\vec{v}_i=d\vec{r}_i/dt$, $\vec{a}_i=d^2\vec{r}_i/dt^2$, $\dot{\vec{a}}_i=d^3\vec{r}_i/dt^3\cdot\cdot\cdot$, we get $^{(3)}$ 
$$\vec{r}¡¯_{ij}(t,t¡¯)\simeq\vec{r}_i(t)-\vec{r}¡¯_j(t-r_{ij}/c)=\vec{r}_i(t)-\vec{r}_j(t)+{{\vec{v}_i(t)}\over{c}}r_{ij}(t)-{{\vec{a}_i(t)}\over{2c^2}}r^2_{ij}(t)+{{\dot{\vec{a}}_i(t)}\over{6c^3}}r^3_{ij}(t)+\cdot\cdot\cdot$$
\begin{equation}
=\vec{r}_{ij}(t)+{{\vec{v}_i(t)}\over{c}}{r}_{ij}(t)-{{\vec{a}_i(t)}\over{2c^2}}r^2_{ij}(t)+{{\dot{\vec{a}}_i(t)}\over{6c^3}}r^3_{ij}(t)+\cdot\cdot\cdot
\end{equation}
\begin{equation}
\vec{v}¡¯_i(t¡¯)\simeq\vec{v}_i(t)-{{\vec{a}_i(t)}\over{c}}r_{ij}(t)+{{\dot{\vec{a}}_i(t)}\over{2c^2}}r^2_{ij}(t)+\cdot\cdot\cdot~~~~~~~\vec{a}¡¯_i(t¡¯)\simeq(t)-{{\dot{\vec{a}}_i(t)}\over{c}}r_{ij}(t)+\cdot\cdot\cdot
\end{equation}
Thus, according to Eq.(16), the retarded electrical dipole moment becomes approximately
\begin{equation}
\vec{P}_i=q_i\vec{r}¡¯_{ij}=q_i(\vec{r}_{ij}+{{\vec{v}_i}\over{c}}r_{ij}-{{\vec{a}_i}\over{2c^2}}r^2_{ij}+{{\dot{\vec{a}}_i}\over{6c^3}}r^3_{ij}+\cdot\cdot\cdot)
\end{equation}
When $t$ is big enough, we have $e^{-2\beta_i{t}}\rightarrow{0}$, the second item on the left side of Eq.(13) can be omitted, so we have $\vec{r}_{ij}=\vec{A}_i\sin(\omega{t}-\theta_0)$ as well as
\begin{equation}
\vec{v}_i=\vec{A}_i\omega\cos(\omega{t}-\theta_{i0})~~~~~~\vec{a}_i=-\vec{A}_i\omega^2\sin(\omega{t}-\theta_{i0})~~~~~~\dot{\vec{a}}_i=-\vec{A}_i\omega^3\cos(\omega{t}-\theta_{i0})
\end{equation}
If there is only one electrical dipole, the directions of $\vec{r}_{ij}$, $\vec{v}_i$, $\vec{a}_i$ and $\dot{\vec{a}}_i$ are on the same straight line. In this case, Eq.(18) can be written as
$$P_i={{q^2_i{E}_0\cos\theta_0}\over{\omega^2}}[\sin(\omega{t}-\theta_{i0})+\alpha_i{E}_0\cos(\omega{t}-\theta_{i0})\mid\sin(\omega{t}-\theta_{i0})\mid$$
\begin{equation}
+{{\alpha^2_i{E}^2_0}\over{2!}}\sin(\omega{t}-\theta_{i0})\mid\sin(\omega{t}-\theta_{i0})\mid^2-{{\alpha^3_i{E}^3_0}\over{3!}}\cos(\omega{t}-\theta_{i0})\mid\sin(\omega{t}-\theta_{i0})\mid^3+\cdot\cdot\cdot]
\end{equation}
Here $\alpha_i=q_i\cos\theta_{i0}/(cm_i\omega)$. In this way, the non-linear relation between polarization strength of ferroelectrics and electrical field strength is introduced automatically. The order of magnitude of non-linear items or the value of parameter $\alpha_i=q_i/(cm_i\omega)$ is estimated below. For ferroelectrics composed of $BaTiO_3$, $m_i=8.02\times{10}^{-26}Kg$ is the mass of $T_i$ iron with $q_i=4e$. Let $\theta_{i0}=0$ for simplification, we can get $\alpha_i=2.66\times{10}^{-2}/\omega$. Taking $\omega=2\pi\times{10}^3$, $E_0=10^3$ $V/M$, we have $\alpha_i{E}_0=0.42$. The non-linear effect is quite strong. Similar, for $O$ iron, $m_i=2.67\times{1.67}\times{10}^{-26}Kg$, $q_i=-2e$, we have $\alpha_i{E}_0=-0.64$. If frequency becomes lower, the non-linear effect would become bigger, for example, taking $\omega=20\pi$, $E_0=10^3V/A$, we have $\alpha_i{E}_0=4.2$ for $Ti$ iron and $\alpha_i{E}_0=64$ for $O$ iron. When frequency becomes high, the non-linear items become small, when frequency become high enough, the non-linear effect would become small enough so that it can¡¯t be observed. The frequency is just so-called cut-off frequency. On the other hand, if the natural frequency of dipole $\omega_0\neq{0}$, we have $\omega_0=10^{14}\sim{10}^{16}>>\omega$. When electrical field is not very strong, for example $E_0=10^6V/M$, we have $\alpha_i{E}_0=10^{-6}\sim{10}^{-8}$, the non-linear effect is too weak to be observed. So the hypotheses of lattice vibrating frozen in soft modal theory is important for us to get observable non-linear relation. It is obvious that all results above coincide with experiments. The parameters of each item in Eq.(20) are $\pm(\alpha_i{E}_0)^n/n!$. Because $\alpha_i{E}_0$ is finite, when $n\rightarrow\infty$ we have $(\alpha_i{E}_0)^n/n!\rightarrow{0}$. Write the $n$ item of the series as $S_n$, we have $S_{n+1}/S_n\sim\alpha_i{E}_0/n$. When $n\rightarrow\infty$ we have $S_{n+1}/S_n\rightarrow{0}$, So the series is convergent. Though when frequency is low and electric field is not very high and $n$ is not very big, we can have $(\alpha_i{E}_0)^n/n!>1$, even $(\alpha_i{E}_0)^n/n!>>1$. In this case many items are needed to engage enough accuracy.
\par
Eq.(20) is based on simple model, but practical ferroelectrics are very complex. For practical ferroelectrics, many factors should be considered, for example, the practical lattice structure, pseudo spin waves and statistical nature of system, interactions between dipoles, as well as interactions between irons and electrons and so on. We can put all of those factors in a nutshell as a factor $k_i$, and write the polarization relation of practical ferroelectrics as
$$P_i={{q^2_i{E}_0\cos\theta_{i0}}\over{m_i\omega^2}}[k_{1i}\sin(\omega{t}-\theta_{i0})+k_{2i}\alpha_i{E}_0\cos(\omega{t}-\theta_{i0})\mid\sin(\omega{t}-\theta_{i0})\mid$$
\begin{equation}
+{{k_{3i}\alpha^2_i{E}^2_0}\over{2!}}\sin(\omega{t}-\theta_{i0})\mid\sin(\omega{t}-\theta_{i0})\mid^2-{{k_{4i}\alpha^3_i{E}^3_0}\over{3!}}\cos(\omega{t}-\theta_{i0})\mid\sin(\omega{t}-\theta_{i0})\mid^3+\cdot\cdot\cdot]
\end{equation}
Suppose there are $N$ effective dipoles in ferroelectrics, the total polarization strength of ferroelectrics in external alternating electrical field is
$$P=\sum^{N}_{i=1}{{q^2_i{E}_0\cos\theta_{i0}}\over{m_i\omega^2}}[k_{1i}\sin(\omega{t}-\theta_{i0})+k_{2i}\alpha_i{E}_0\cos(\omega{t}-\theta_{i0})\mid\sin(\omega{t}-\theta_{i0})\mid$$
\begin{equation}
+{{k_{3i}\alpha^2_i{E}^2_0}\over{2!}}\sin^3(\omega{t}-\theta_{i0})-{{k_{4i}\alpha^3_i{E}^3_0}\over{3!}}\cos(\omega{t}-\theta_{i0})\mid\sin(\omega{t}-\theta_{i0})\mid^3+\cdot\cdot\cdot]
\end{equation}
The first item can be written as
\begin{equation}
\sum^{N}_{i=1}{{q^2_i{E}_0{k}_{1i}\cos\theta_{i0}}\over{m_i\omega^2}}\sin(\omega{t}-\theta_{i0})=\varepsilon_0(x_1\sin\omega{t}-y_1\cos\omega{t})
\end{equation}
\begin{equation}
x_1=\sum^{N}_{i=1}{{q^2_i{E}_0{k}_{1i}\cos^2\theta_{i0}}\over{\varepsilon_0{m}_i\omega^2}}~~~~~~~~~~~~y_1=\sum^{N}_{i=1}{{q^2_i{E}_0{k}_{1i}\sin\theta_{i0}\cos\theta_{i0}}\over{\varepsilon_0{m}_i\omega^2}}
\end{equation}
In the formula, $x_1$ and $y_1$ can be regarded independent each other, so we regard them as the ordinates of the points in a plane. Let $\chi_{e1}=\sqrt{x^2_1+y^2_1}$, $y_1/x_1=tg\theta_1$, we can always write $x_1$ and $y_1$ as $x_1=\chi_{e1}\cos\theta_1$ and $y_1=\chi_{e1}\sin\theta_1$. So Eq.(23) can be re-written as
\begin{equation}
\sum^{N}_{i=1}{{q^2_i{E}_0{k}_{1i}\cos\theta_{i0}}\over{m_i\omega^2}}\sin(\omega{t}-\theta_{i0})=\varepsilon_0\chi_{e1}\sin(\omega{t}-\theta_1)
\end{equation}
Similarly, the second item in Eq.(22) can be written as
$$\sum^{N}_{i=1}{{q^2_i{k}_{2i}\alpha_i{E}^2_0\cos\theta_{i0}}\over{m_i\omega^2}}\cos(\omega{t}-\theta_{i0})\mid\sin(\omega{t}-\theta_{i0})\mid$$
\begin{equation}
=\varepsilon_0(x_2\sin\omega{t}\cos\omega{t}-y_2\cos\omega{t}\sin\omega{t}+z_2\sin^2\omega{t}-z¡¯_2\sin^2\omega{t})
\end{equation}
\begin{equation}
x_2=\sum^{N}_{j=1}{{q^2_i{k}_{2i}\alpha_i{E}^2_0\cos^2\theta_{i0}\mid\cos\theta_{i0}\mid}\over{\varepsilon_0{m}_i\omega^2}}~~~~~~~~~y_2=\sum^{N}_{i=1}{{q^2_i{k}_{2i}\alpha_i{E}^2_0\sin\theta_{i0}\cos\theta_{i0}\mid\sin\theta_{i0}\mid}\over{\varepsilon_0{m}_i\omega^2}}
\end{equation}
\begin{equation}
z_2=\sum^{N}_{i=1}{{q^2_i{k}_{2i}\alpha_i{E}^2_0\sin\theta_{i0}\cos\theta_{i0}\mid\cos\theta_{i0}\mid}\over{\varepsilon_0{m}_i\omega^2}}~~~~~~~~~z¡¯_2=\sum^{N}_{j=1}{{q^2_i{k}_{2i}\alpha_i{E}^2_0\cos^2\theta_{i0}\mid\sin\theta_{i0}\mid}\over{\varepsilon_0{m}_i\omega^2}}
\end{equation}
Because of $\sin\theta_{i0}=\cos(\pi/2-\theta_{i0})$, it can be considered to be the same to take sum over $\sin\theta_{i0}\mid\cos\theta_{i0}\mid$ and over $\cos\theta_{i0}\mid\sin\theta_{i0}\mid$. So we have $z_2=z¡¯_2$. In this way, there are three independent parameters on the right side of Eq.(26). We can take them as $\chi_2$, $\theta_2$, $\theta_3$, and write Eq.(26) as 
\begin{equation}
\sum^{N}_{i=1}{{q^2_i{k}_{2i}\alpha_i{E}^2_0\cos\theta_{i0}}\over{m_i\omega^2}}\cos(\omega{t}-\theta_{i0})\mid\sin(\omega{t}-\theta_{i0})\mid=\varepsilon_0\chi_{e2}{E}^2_0\cos(\omega{t}-\theta_2)\mid\sin(\omega{t}-\theta_3)\mid
\end{equation}
Similarly, the third item has four independent parameters $\chi_3$, $\theta_4$, $\theta_5$, $\theta_6$, and the third item has five independent parameters $\chi_4$, $\theta_7$, $theta_8$, $\theta_9$, $\theta_{10}$, and so on. At last, Eq.(22) can be written as
$$P=\varepsilon_0{E}_0[\chi_{e1}\sin(\omega{t}-\theta_1)+\chi_{e2}{E}_0\cos(\omega{t}-\theta_2)\mid\sin(\omega{t}-\theta_3)\mid$$
$$+\chi_{e3}{E}^2_0\sin(\omega{t}-\theta_4)\mid\sin(\omega{t}-\theta_5)\sin(\omega{t}-\theta_6)$$
\begin{equation}
+\chi_{e4}{E}^3_0\cos(\omega{t}-\theta_7)\mid\sin(\omega{t}-theta_8)\sin(\omega{t}-\theta_9)\sin(\omega{t}-\theta_{10})\mid+\cdot\cdot\cdot]
\end{equation}
The parameters $\chi_{en}$ and $\theta_n$ are different for different ferroelectrics and can be determined by experiments. The formula can also be written as
$$P=\varepsilon_0\chi_{e1}{E}_0[\sin(\omega{t}-\theta_1)+b_1\cos(\omega{t}-\theta_2)\mid\sin(\omega{t}-\theta_3)\mid$$
$$+b_2\sin(\omega{t}-\theta_4)\mid\sin(\omega{t}-\theta_5)\sin(\omega{t}-\theta_6)\mid$$
\begin{equation}
+b_3\cos(\omega{t}-\theta_7)\mid\sin(\omega{t}-\theta_8)\sin(\omega{t}-\theta_9)\sin(\omega{t}-\theta_{10})\mid+\cdot\cdot\cdot]
\end{equation}
Here $b_n=\chi_{en-1}{E}^{n-1}_0/\chi_{e1}$ with $b_n\rightarrow{0}$ when $n\rightarrow\infty$. Then, we consider the simplest situation to take two items in (31). Let $b_1=0.1$, $b_n=0(n>1)$ and $\theta_1=\theta_2=\theta_3=\theta$. When $\omega{t}=\theta$ we get $P=0$. In this case $\theta$ is just the polarization hysteretic phase. Take $\theta=30^0$ and $60^0$. Tate $E_0$ as the unit of abscissa, let $P_0$ represent the maximum value of $P$ and take $P_0/(\varepsilon_0\chi_{e1})$ as the unit of ordinate, we get the hysteretic loops shown in Fig.5 and 6. They coincide with practical results in the medium frequency fields.

\begin{figure}[h]
\begin{center}
\begin{picture}(200,240)(-100,-120)
\put(-100,-120){\tiny \grid(200,240)(50,20)[-100,-120]}
\drawline(0,-46)(17,-31)(34,-15)(50,0)(64,19)(77,37)(87,54)(94,69)(98,82)(100,91)(98,97)(94,100)
(87,100)(77,96)(64,91)(50,83)(34,72)(17,59)(0,46)(-17,31)
(-34,15)(-50,0)(-64,-19)(-77,-37)(-87,-54)(-94,-69)(-98,-82)
(-100,-91)(-98,-97)(-94,-100)(-87,-100)(-77,-96)(-64,-91)(-50,-83)(-34,-72)(-17,-59)(0,-46)
\dottedline{1.4}(0,0)(17,11)(34,24)(50,37)(64,51)(77,67)(87,79)(94,93)(98,102)
(100,108)(98,114)(94,115)(87,113)
\end{picture}
\end{center}
\caption[]{}
\end{figure}

\begin{figure}[h]
\begin{center}
\begin{picture}(200,240)(-100,-120)
\put(-100,-120){\tiny \grid(200,240)(50,20)[-100,-120]}
\drawline(0,-83)(17,-72)(34,-59)(50,-46)(64,-31)(77,-15)(87,0)(94,19)(98,37)(100,54)(98,69)(94,82)
(87,91)(77,97)(64,100)(50,100)(34,96)(17,91)(0,83)(-17,72)
(-34,59)(-50,46)(-64,31)(-77,15)(-85,0)(-94,-19)(-98,-37)
(-100,-54)(-98,-69)(-94,-82)(-87,-91)(-77,-97)(-64,-100)(-50,-100)(-34,-96)(-17,-91)(0,-83)
\dottedline{1.4}(0,0)(17,7)(34,18)(50,29)(64,43)(77,60)(87,76)(94,93)(98,108)(100,131)
\end{picture}
\end{center}
\caption[]{}
\end{figure}

For low frequencies fields, we stochastically take $b_1=0.400$, $b_2=0.256$, $b_3=0.461$, $b_4=0.164$, $b_5=0.665$, $b_6=0.184$, $b_7=-0.419$, $b_8=0.0420$, $b_9=0.034$, $b_{10}=0.027$, $b_{n>10}=0$, $\theta_1=0.300$, $\theta_2=2.100$, $\theta_3=1.500$, $\theta_4=-0.300$, $\theta_5=1.300$, $\theta_6=6.000$, $\theta_7=6.000$, $\theta_8=1.200$, $\theta_{n>8}=1.050+(2n)mod6$, get Fig.7. Similarly, take $b_1=0.464$, $b_2=0.147$, $b_3=0.067$, $b_4=0.164$, $b_5=0.475$, $b_6=0.021$, $b_7=-0.042$, $b_8=0.042$, $b_9=0.034$, $b_{10}=0.027$, $b_{n>10}=0$, $\theta_1=0.270$, $\theta_2=3.180$, $\theta_3=1.780$, $\theta_4=0.080$, $\theta_5=0.160$, $\theta_6=3.030$, $\theta_7=5.580$, $\theta_8=5.560$, $\theta_{n>8}=1+0.200n$, get Fig.8. Two of them are good rectangles, showing that Eq.(31) can also be used to describe the hysteretic loops in low frequency fields.

\begin{figure}[h]
\begin{center}
\begin{picture}(200,240)(-100,-120)
\put(-100,-120){\tiny \grid(200,240)(50,20)[-100,-120]}
\drawline(0,-84)(1,-83)(2,-82)(3,-81)(5,-79)(6,-78)(7,-77)(8,-75)(10,-74)(11,-72)
(12,-71)(13,-69)(15,-67)(16,-65)(17,-63)(18,-61)(19,-59)(21,-57)(22,-55)(23,-53)
(24,-51)(26,-48)(27,-46)(28,-44)(29,-41)(30,-39)(32,-36)(33,-34)(34,-31)(35,-29)
(36,-26)(37,-23)(39,-21)(40,-18)(41,-15)(42,-12)(43,-10)(44,-7)(45,-4)(47,-1)
(48,0)(49,3)(50,6)(51,9)(52,11)(53,14)(54,17)(55,19)(56,22)(57,25)
(58,27)(59,30)(60,32)(61,35)(62,37)(63,40)(64,42)(65,44)(66,47)(67,49)
(68,51)(69,53)(70,55)(71,57)(72,59)(72,61)(73,62)(74,64)(75,66)(76,68)
(77,69)(77,71)(78,73)(79,74)(80,76)(80,77)(81,78)(82,80)(83,81)(83,82)
(84,83)(85,85)(85,86)(86,87)(87,88)(87,89)(88,90)(88,90)(89,91)(89,92)
(90,93)(91,93)(91,94)(92,94)(92,95)(92,95)(93,96)(93,96)(94,96)(94,97)
(95,97)(95,97)(95,97)(96,97)(96,97)(96,98)(97,98)(97,98)(97,99)(97,99)
(98,99)(98,99)(98,99)(98,99)(99,99)(99,99)(99,99)(99,99)(99,98)(99,98)
(99,98)(99,99)(99,99)(99,99)(99,100)(100,100)(99,100)(99,100)(99,101)(99,101)
(99,101)(99,101)(99,101)(99,101)(99,101)(99,101)(99,102)(98,102)(98,102)(98,102)
(98,102)(97,101)(97,101)(97,101)(97,101)(96,101)(96,101)(96,101)(95,101)(95,101)
(95,101)(94,101)(94,101)(93,101)(93,101)(92,101)(92,101)(92,101)(91,100)(91,100)
(90,100)(89,100)(89,100)(88,100)(88,100)(87,100)(87,100)(86,100)(85,100)(85,100)
(84,100)(83,100)(83,100)(82,100)(81,100)(80,100)(80,100)(79,100)(78,100)(77,100)
(77,100)(76,100)(75,100)(74,100)(73,100)(72,100)(72,99)(71,99)(70,99)(69,99)
(68,99)(67,99)(66,99)(65,99)(64,99)(63,98)(62,98)(61,98)(60,98)(59,98)
(58,98)(57,98)(56,97)(55,97)(54,97)(53,97)(52,97)(51,97)(50,97)(49,96)
(48,96)(47,96)(45,96)(44,96)(43,96)(42,96)(41,96)(40,95)(39,95)(37,95)
(36,95)(35,95)(34,95)(33,95)(32,95)(30,95)(29,95)(28,95)(27,94)(26,94)
(24,94)(23,93)(22,93)(21,93)(19,93)(18,92)(17,92)(16,92)(15,91)(13,91)
(12,91)(11,90)(10,90)(8,90)(7,89)(6,88)(5,88)(3,87)(2,86)(1,85)
(0,84)(-1,83)(-2,82)(-3,81)(-5,79)(-6,78)(-7,77)(-8,75)(-10,74)(-11,72)
(-12,71)(-13,69)(-15,67)(-16,65)(-17,63)(-18,61)(-19,59)(-21,57)(-22,55)(-23,53)
(-24,51)(-26,48)(-27,46)(-28,44)(-29,41)(-30,39)(-32,36)(-33,34)(-34,31)(-35,29)
(-36,26)(-37,23)(-39,21)(-40,18)(-41,15)(-42,12)(-43,10)(-44,7)(-45,4)(-47,1)
(-48,0)(-49,-3)(-50,-6)(-51,-9)(-52,-11)(-53,-14)(-54,-17)(-55,-19)(-56,-22)(-57,-25)
(-58,-27)(-59,-30)(-60,-32)(-61,-35)(-62,-37)(-63,-40)(-64,-42)(-65,-44)(-66,-47)(-67,-49)
(-68,-51)(-69,-53)(-70,-55)(-71,-57)(-72,-59)(-72,-61)(-73,-62)(-74,-64)(-75,-66)(-76,-68)
(-77,-69)(-77,-71)(-78,-73)(-79,-74)(-80,-76)(-80,-77)(-81,-78)(-82,-80)(-83,-81)(-83,-82)
(-84,-83)(-85,-85)(-85,-86)(-86,-87)(-87,-88)(-87,-89)(-88,-90)(-88,-90)(-89,-91)(-89,-92)
(-90,-93)(-91,-93)(-91,-94)(-92,-94)(-92,-95)(-92,-95)(-93,-96)(-93,-96)(-94,-96)(-94,-97)
(-95,-97)(-95,-97)(-95,-97)(-96,-97)(-96,-97)(-96,-98)(-97,-98)(-97,-98)(-97,-99)(-97,-99)
(-98,-99)(-98,-99)(-98,-99)(-98,-99)(-99,-99)(-99,-99)(-99,-99)(-99,-99)(-99,-98)(-99,-98)
(-99,-98)(-99,-99)(-99,-99)(-99,-99)(-99,-100)(-100,-100)(-99,-100)(-99,-100)(-99,-101)(-99,-101)
(-99,-101)(-99,-101)(-99,-101)(-99,-101)(-99,-101)(-99,-101)(-99,-102)(-98,-102)(-98,-102)(-98,-102)
(-98,-102)(-97,-101)(-97,-101)(-97,-101)(-97,-101)(-96,-101)(-96,-101)(-96,-101)(-95,-101)(-95,-101)
(-95,-101)(-94,-101)(-94,-101)(-93,-101)(-93,-101)(-92,-101)(-92,-101)(-92,-101)(-91,-100)(-91,-100)
(-90,-100)(-89,-100)(-89,-100)(-88,-100)(-88,-100)(-87,-100)(-87,-100)(-86,-100)(-85,-100)(-85,-100)
(-84,-100)(-83,-100)(-83,-100)(-82,-100)(-81,-100)(-80,-100)(-80,-100)(-79,-100)(-78,-100)(-77,-100)
(-77,-100)(-76,-100)(-75,-100)(-74,-100)(-73,-100)(-72,-100)(-72,-99)(-71,-99)(-70,-99)(-69,-99)
(-68,-99)(-67,-99)(-66,-99)(-65,-99)(-64,-99)(-63,-98)(-62,-98)(-61,-98)(-60,-98)(-59,-98)
(-58,-98)(-57,-98)(-56,-97)(-55,-97)(-54,-97)(-53,-97)(-52,-97)(-51,-97)(-50,-97)(-49,-96)
(-48,-96)(-47,-96)(-45,-96)(-44,-96)(-43,-96)(-42,-96)(-41,-96)(-40,-95)(-39,-95)(-37,-95)
(-36,-95)(-35,-95)(-34,-95)(-33,-95)(-32,-95)(-30,-95)(-29,-95)(-28,-95)(-27,-94)(-26,-94)
(-24,-94)(-23,-93)(-22,-93)(-21,-93)(-19,-93)(-18,-92)(-17,-92)(-16,-92)(-15,-91)(-13,-91)
(-12,-91)(-11,-90)(-10,-90)(-8,-90)(-7,-89)(-6,-88)(-5,-88)(-3,-87)(-2,-86)(-1,-85)
\end{picture}
\end{center}
\caption[]{}
\end{figure}

\begin{figure}[h]
\begin{center}
\begin{picture}(200,240)(-100,-120)
\put(-100,-120){\tiny \grid(200,240)(50,20)[-100,-120]}
\drawline(0,-92)(1,-91)(2,-91)(3,-90)(5,-89)(6,-89)(7,-88)(8,-87)(10,-86)(11,-85)
(12,-84)(13,-83)(15,-82)(16,-81)(17,-80)(18,-79)(19,-78)(21,-76)(22,-75)(23,-74)
(24,-72)(26,-71)(27,-69)(28,-68)(29,-66)(30,-64)(32,-63)(33,-61)(34,-59)(35,-57)
(36,-56)(37,-54)(39,-52)(40,-50)(41,-48)(42,-46)(43,-43)(44,-41)(45,-39)(47,-37)
(48,-35)(49,-33)(50,-30)(51,-28)(52,-26)(53,-23)(54,-21)(55,-19)(56,-16)(57,-14)
(58,-12)(59,-9)(60,-7)(61,-4)(62,-2)(63,0)(64,2)(65,4)(66,6)(67,9)
(68,11)(69,13)(70,16)(71,18)(72,20)(72,22)(73,25)(74,27)(75,29)(76,31)
(77,33)(77,36)(78,38)(79,40)(80,42)(80,44)(81,46)(82,48)(83,50)(83,51)
(84,53)(85,55)(85,57)(86,59)(87,60)(87,62)(88,64)(88,65)(89,67)(89,68)
(90,70)(91,71)(91,73)(92,74)(92,75)(92,77)(93,78)(93,79)(94,81)(94,82)
(95,83)(95,84)(95,85)(96,86)(96,87)(96,88)(97,89)(97,90)(97,91)(97,92)
(98,92)(98,93)(98,94)(98,95)(99,95)(99,96)(99,97)(99,97)(99,98)(99,98)
(99,99)(99,99)(99,100)(99,100)(99,101)(100,101)(99,101)(99,101)(99,102)(99,102)
(99,102)(99,102)(99,102)(99,102)(99,102)(99,102)(99,102)(98,102)(98,102)(98,102)
(98,102)(97,102)(97,101)(97,101)(97,101)(96,101)(96,101)(96,101)(95,101)(95,101)
(95,101)(94,101)(94,101)(93,101)(93,101)(92,101)(92,101)(92,101)(91,101)(91,101)
(90,101)(89,101)(89,101)(88,101)(88,101)(87,101)(87,101)(86,101)(85,101)(85,101)
(84,101)(83,101)(83,101)(82,101)(81,101)(80,101)(80,101)(79,101)(78,101)(77,101)
(77,101)(76,101)(75,101)(74,101)(73,101)(72,101)(72,101)(71,101)(70,101)(69,101)
(68,100)(67,100)(66,100)(65,100)(64,100)(63,100)(62,100)(61,100)(60,100)(59,100)
(58,99)(57,99)(56,99)(55,99)(54,99)(53,99)(52,99)(51,99)(50,99)(49,99)
(48,99)(47,99)(45,99)(44,99)(43,98)(42,98)(41,98)(40,98)(39,98)(37,98)
(36,98)(35,98)(34,98)(33,98)(32,98)(30,98)(29,98)(28,98)(27,98)(26,98)
(24,97)(23,97)(22,97)(21,97)(19,97)(18,97)(17,97)(16,97)(15,96)(13,96)
(12,96)(11,96)(10,95)(8,95)(7,95)(6,94)(5,94)(3,93)(2,93)(1,93)
(0,92)(-1,91)(-2,91)(-3,90)(-5,89)(-6,89)(-7,88)(-8,87)(-10,86)(-11,85)
(-12,84)(-13,83)(-15,82)(-16,81)(-17,80)(-18,79)(-19,78)(-21,76)(-22,75)(-23,74)
(-24,72)(-26,71)(-27,69)(-28,68)(-29,66)(-30,64)(-32,63)(-33,61)(-34,59)(-35,57)
(-36,56)(-37,54)(-39,52)(-40,50)(-41,48)(-42,46)(-43,43)(-44,41)(-45,39)(-47,37)
(-48,35)(-49,33)(-50,30)(-51,28)(-52,26)(-53,23)(-54,21)(-55,19)(-56,16)(-57,14)
(-58,12)(-59,9)(-60,7)(-61,4)(-62,2)(-63,0)(-64,-2)(-65,-4)(-66,-6)(-67,-9)
(-68,-11)(-69,-13)(-70,-16)(-71,-18)(-72,-20)(-72,-22)(-73,-25)(-74,-27)(-75,-29)(-76,-31)
(-77,-33)(-77,-36)(-78,-38)(-79,-40)(-80,-42)(-80,-44)(-81,-46)(-82,-48)(-83,-50)(-83,-51)
(-84,-53)(-85,-55)(-85,-57)(-86,-59)(-87,-60)(-87,-62)(-88,-64)(-88,-65)(-89,-67)(-89,-68)
(-90,-70)(-91,-71)(-91,-73)(-92,-74)(-92,-75)(-92,-77)(-93,-78)(-93,-79)(-94,-81)(-94,-82)
(-95,-83)(-95,-84)(-95,-85)(-96,-86)(-96,-87)(-96,-88)(-97,-89)(-97,-90)(-97,-91)(-97,-92)
(-98,-92)(-98,-93)(-98,-94)(-98,-95)(-99,-95)(-99,-96)(-99,-97)(-99,-97)(-99,-98)(-99,-98)
(-99,-99)(-99,-99)(-99,-100)(-99,-100)(-99,-101)(-100,-101)(-99,-101)(-99,-101)(-99,-102)(-99,-102)
(-99,-102)(-99,-102)(-99,-102)(-99,-102)(-99,-102)(-99,-102)(-99,-102)(-98,-102)(-98,-102)(-98,-102)
(-98,-102)(-97,-102)(-97,-101)(-97,-101)(-97,-101)(-96,-101)(-96,-101)(-96,-101)(-95,-101)(-95,-101)
(-95,-101)(-94,-101)(-94,-101)(-93,-101)(-93,-101)(-92,-101)(-92,-101)(-92,-101)(-91,-101)(-91,-101)
(-90,-101)(-89,-101)(-89,-101)(-88,-101)(-88,-101)(-87,-101)(-87,-101)(-86,-101)(-85,-101)(-85,-101)
(-84,-101)(-83,-101)(-83,-101)(-82,-101)(-81,-101)(-80,-101)(-80,-101)(-79,-101)(-78,-101)(-77,-101)
(-77,-101)(-76,-101)(-75,-101)(-74,-101)(-73,-101)(-72,-101)(-72,-101)(-71,-101)(-70,-101)(-69,-101)
(-68,-100)(-67,-100)(-66,-100)(-65,-100)(-64,-100)(-63,-100)(-62,-100)(-61,-100)(-60,-100)(-59,-100)
(-58,-99)(-57,-99)(-56,-99)(-55,-99)(-54,-99)(-53,-99)(-52,-99)(-51,-99)(-50,-99)(-49,-99)
(-48,-99)(-47,-99)(-45,-99)(-44,-99)(-43,-98)(-42,-98)(-41,-98)(-40,-98)(-39,-98)(-37,-98)
(-36,-98)(-35,-98)(-34,-98)(-33,-98)(-32,-98)(-30,-98)(-29,-98)(-28,-98)(-27,-98)(-26,-98)
(-24,-97)(-23,-97)(-22,-97)(-21,-97)(-19,-97)(-18,-97)(-17,-97)(-16,-97)(-15,-96)(-13,-96)
(-12,-96)(-11,-96)(-10,-95)(-8,-95)(-7,-95)(-6,-94)(-5,-94)(-3,-93)(-2,-93)(-1,-93)

\end{picture}
\end{center}
\caption[]{}
\end{figure}

\clearpage

\par
As for polarization curves, by using Eq.(12), Eq.(22) can be written as
$$P=\sum^{N}_{i=1}{{q^2_i{k}_{1i}{E}_0\cos\theta_{i0}}\over{m_i\omega^2}}[\sin(\omega{t}-\theta_{i0})-{{\cos^2_{i0}}\over{\sin\theta_{i0}}}e^{-2\beta_i{t}}]$$
\begin{equation}
+\sum^{N}_{i=1}{{q^2_i{k}_{2i}\alpha_i{E}^2_0\cos\theta_{i0}}\over{m_i{c}\omega^2}}[\cos(\omega{t}-\theta_{i0})-\cos\theta_{i0}e^{-2\beta_i{t}}]\mid\sin(\omega{t}-\theta_{i0})-{{\cos^2_{i0}}\over{\sin\theta_{i0}}}e^{-2\beta_i{t}}\mid\cdot\cdot\cdot
\end{equation}
Similar to Eq.(30), by considering relation $2\beta_i{t}=\mid{t}g\theta_{i0}\mid\omega{t}$, the formula above can be written as 
$$P=\varepsilon_0\chi_{e1}{E}_0\{\sin(\omega{t}-\theta_1)+g_1{e}^{-k_1\mid{t}g\theta_1\mid\omega{t}}$$
\begin{equation}
+b_1[\cos(\omega{t}-\theta_2)+g_2{e}^{-k_2\mid{t}g\theta_2\mid\omega{t}}]\mid\sin(\omega{t}-\theta_3)+g_3{e}^{-k_3\mid{t}g\theta_2\mid\omega{t}}\mid\}\cdot\cdot\cdot
\end{equation}
Though each electrical domain is polarized, whole ferroelectrics are medium generally at beginning. The initial condition is $P=0$ when $t=0$. Only taking first two items, we have $g_1=\sin\theta_1$, $g_3=\sin\theta_3$ and get at last
$$P=\varepsilon_0\chi_{e1}{E}_0\{\sin(\omega{t}-\theta_1)+\sin\theta_1{e}^{-k_1\mid{t}g\theta_1\mid\omega{t}}$$
\begin{equation}
+b_1[\cos(\omega{t}-\theta_2)+g_2{e}^{-k_2\mid{t}g\theta_2\mid\omega{t}}]\mid\sin(\omega{t}-\theta_3)+\sin\theta_3{e}^{-k_3\mid{t}g\theta_2\mid\omega{t}}\mid\}
\end{equation}
The parameters $k_i>0$, the value should be decided by experiments. Taking $\theta_1=\theta_2=\theta_3=\theta=30^0$, $b=0.1$, $k_1=k_2=k_3=1$, $g_2=1$, we get polarization curve shown in Fig.5. Taking $\theta=60^0$, $b=0.1$, $k_1=k_2=k_3=0.1$, $g_2=-\cos{60}^0=-0.5$, we get polarization curve shown in Fig.6.
\par
The results are the same for the magnetic hysteretic loops and magnetization curves of ferromagnetics. As we known that magnetic phenomena can be equivalently described by the concept of magnetic charges. According to quantum mechanics, there exist the exchange forces between charges. When magnetic moments in magnetic domains are arranged in parallel, the system¡¯s energy is lowest so that the system is most stable. Magnetic moments are caused when electrons move around atomic nucleuses. If there is no external field, electron¡¯s motions are stationary and the magnetic moments of atoms are also stationary without varying with time. The situation is similar to the soft modal theory of ferroelectrics. So we can thinks that the vibration of magnetic dipole is frozen with nature frequency $\omega_0=0$ when external fields do not exist. In this way, Eq.(11) can also be used to described the motion of magnetic dipole in alternating magnetic field. Similar to Eq.(20), in the alternating magnetic field $\vec{H}=\vec{H}_0\sin\omega{t}$, the magnetic moment of a single magnetic dipole is
$$M_i={{q^2_i\cos\theta_{i0}{H}_0}\over{m_i\omega^2}}[\sin(\omega{t}-\theta_{i0})+\alpha_i{H}_0\cos(\omega{t}-\theta_{i0})\mid\sin(\omega{t}-\theta_{i0})\mid$$
\begin{equation}
+{{\alpha^2_i{H}^2_0}\over{2!}}\sin(\omega{t}-\theta_{i0})\mid\sin(\omega{t}-\theta_{i0})\mid^2-{{\alpha^3_i{H}^3_0}\over{3!}}\cos(\omega{t}-\theta_{i0})\mid\sin(\omega{t}-\theta_{i0})\mid^3+\cdot\cdot\cdot]
\end{equation}
The order of magnitude of parameter $\alpha_i=q_i/(cm_i\omega)$ is estimated below. It is known that only out-shell electrons have contribution to magnetic moments. Suppose there is only one out-shell electron in an atom, each atom has a magnetic moment called Bohr magneton $M_i=\mu_0{e}\hbar/(2m_3)\simeq{1.2}\times{10}^{-29}Wb\cdot{M}$. Suppose the distance between atoms is about $r_i\simeq{10}^{-10}N\cdot{M}/A$, according to the definition of magnetic moment, $M_i=r_i{q}_i$, we have the value of equivalent magnetic charge $q_i\simeq{1.2}\times{10}^{-19}W$. The value is similar to electron¡¯s charge. If atomic nucleus or iron is considered at static and electron moves, $m_i$ should be electron¡¯s mass, we have $\alpha_i=4.4\times{10}^2/\omega$, much big than that of ferroelectrics. Taking $\omega=2\pi\times{10}^2$, $H_0=0.1A/M$, we have $\alpha_i{H}_0=0.7$ with strong non-linearity. If $m_i$ is regarded as the masses of atomic nucleus or iron, the order of magnitude of $\alpha_i$ is similar to that of ferroelectrics. In practical situations, effective masses will be taken. Similar to Eq.(30), we have
$$M=\mu_0{H}_0[\mu_1\sin(\omega{t}-\theta_1)+\mu_2{H}_0\cos(\omega{t}-\theta_2)\mid\sin(\omega{t}-\theta_3)\mid$$
$$+\mu_3{H}^2_0\sin(\omega{t}-\theta_4)\mid\sin(\omega{t}-\theta_5)\sin(\omega{t}-\theta_6)\mid$$
\begin{equation}
+\mu_4{H}^3_0\cos(\omega{t}-\theta_7)\mid\sin(\omega{t}-\theta_8)\sin(\omega{t}-\theta_9)\sin(\omega{t}-\theta_{10})\mid+\cdot\cdot\cdot]
\end{equation}
The magnetic hysteretic loops can also be described well by the formula above. We can also get similar formula to describe magnetization curves of ferromagnetics, but discuss no any more here.
\par
The same method can be used to describer the non-linear phenomena in optics. In the current non-linear optics, interaction between light and materials is described by half-classical method, i.e., to use classical theory of electromagnetic field to describe light¡¯s motion and quantum theory to describe interaction between photons and material particles. The non-linear phenomena can be dealt with well by the non-linear polarization formula $^{(4)}$ 
\begin{equation}
\vec{P}=\varepsilon_0(\chi_{e1}\vec{E}+\chi_{e2}\cdot\vec{E}\vec{E}+\vec{\vec{\chi}}_{e3}:\vec{E}\vec{E}\vec{E}\cdot\cdot\cdot)
\end{equation}
\par
Because optical crystal is not ferroelectrics in general, we regard atoms and molecular as dipoles with nature frequency $\omega_0\neq{0}$. The motion equation and solution of dipoles are Eq.(8) and (9). When $t\rightarrow\infty$, the polarization can also be described by Eq.(30). Taking $\theta_1=\theta_3=\theta_5=\theta_6=\theta_7=\theta_8=\theta_9=\theta_{10}=\theta$, $theta_2=\theta_7=\theta+3\pi/2$, in spite of the phase difference $\theta$ thought it exist actually, we get
\begin{equation}
\vec{P}=\varepsilon_0(\chi_{e1}\vec{E}+\chi_{e2}\cdot{E}\vec{E}+\chi_{e3}{E}^2\vec{E}\cdot\cdot\cdot)
\end{equation}
The result is similar to Eq.(37), from it most of non-linear phenomena just as multiple frequencies, sum frequencies and different frequencies and so on will be described. If non-isotropy of crystal is considered, we should write (38) as (37). Tt should be noted that Eq.(37) is only a special situation of Eq.(30).
\par
Because nuclear mass is much big than electron¡¯s mass, the displacement of electron caused by external field is much big than nuclear one, so we can think that nucleus is at rest in center and electron vibrates around center. Suppose the wavelength of incident light is $\lambda=4\times{10}^{-7}M$, $m_i$ is electron mass, we have $\alpha_i=q_i/(cm_i\omega)\simeq{1.2}\times{10}^{-13}M/V$ The strength of electric field of laser used in non-linear optics is about $E_0\sim{10}^{10}V/M$ in general, so $\alpha_i{E}_0\sim{10}^{-3}$£¬means that the radio between the first item and second item in Eq.(38) is about $10^{-3}$. The value coincides with experiments with $\alpha_i=10^{-12}\sim{10}^{-13}$, $\alpha_i{E}_0=10^{-3}-10^{-4(5)}$. If the light used is not laser with $E_0<{10}^{7}V/M$ in general, we have $\alpha_i{E}_0<{10}^{-6}$, the non-linear effects are too weak to be observed.
\par
In order to explain the origin of non-linear phenomena in optics and get Eq.(37) in theory, the non-linear oscillator model is used at present. The motion equation of oscillator is written as $^{(4)}$ 
\begin{equation}
{{d^2r}\over{dt^2}}+2\beta{{dr}\over{dt}}+\omega^2_0{r}+Dr^2={{F}\over{m}}
\end{equation}
In the formula, $mDr^2$ is a non-linear force to cause non-linear effects. However, it can be seen that the origin of this item is still unclear. Where it comes from? Why does it not taking the form such as $mDr^{1/2}$ or $mDr^3$ and so on? The current theory can¡¯t provide clear explanation. Meanwhile, the value of parameter $D$ can¡¯t be decided by theory now. Quantum mechanics provides a method of perturbation approximation to calculate polarizability, but can¡¯t yet explain the origin of non-linear phenomena. It is obvious that by introducing retarded electromagnetic interaction, we can explain the origin of non-linear effects in ferroelectrics, ferromagnetics and optics well in a united form from the macro-angle of classical electromagnetic theory. The result also prompts us to consider retarded interaction in quantum theory to get more accurate description of ferroelectrics, ferromagnetics, non-linear optics and other fields.
\par
On the other hand, it can be seen that Eq.(14) can not keep unchanged under time reversal with $\vec{v}¡¯_i\rightarrow-\vec{v}¡¯_i$, meaning that retarded electromagnetic interaction is asymmetry under time reversal. It is the same in Eq.(16) under time reversal with $\vec{v}_i\rightarrow-\vec{v}_i$, $\vec{a}_i\rightarrow\vec{a}_i$, $\dot{\vec{a}}_i\rightarrow-\dot{\vec{a}}_i$, meaning that $\vec{r}¡¯$ also violates symmetry of time reversal. So in the non-liner processes of ferroelectrics, ferromagnetics and non-linear optics, symmetry of time reversal is violated obviously. In fact, the polarization and magnetization processes that electromagnetic fields change along positive directions are just the time reversal processes in which the fields change along negative directions. The hysteresis loop¡¯s shapes are similar to that in the cycle processes of heat engines, both are irreversible to produce dissipative heat. For the polarization and magnetization processes of ferroelectrics and ferromagnetics, only electromagnetic interaction is involved. So the irreversibility is caused only by electromagnetic interaction itself, or speaking strictly, by retarded electromagnetic interaction. It is impossible to explain this kind of irreversibility by the theories of coarse graining and mixing current and so on as advocated in current statistical mechanics.

\end{document}